 \documentclass[final,5p,times,twocolumn]{elsarticle}

\usepackage{graphicx}

\usepackage{amssymb}
 \usepackage{amsthm}

\journal{Physics Letters B}

\begin{document}

\begin{frontmatter}



\title{Measurement of positronium hyperfine splitting with quantum oscillation}


\author[icepp]{Y.~Sasaki}
\author[icepp]{A.~Miyazaki}
\author[icepp]{A.~Ishida}
\author[icepp]{T.~Namba}
\author[icepp]{S.~Asai}
\author[icepp]{T.~Kobayashi}
\author[komaba]{H.~Saito}
\author[KEK]{K.~Tanaka}
\author[KEK]{and A.~Yamamoto}

\address[icepp]{Department of Physics, Graduate School of Science, 
and International Center for Elementary Particle Physics,

The University of Tokyo, 7-3-1 Hongo, Bunkyo-ku, Tokyo, 113-0033, Japan}

\address[komaba]{
Institute of Physics,
Graduate School of Arts and Sciences,
The University of Tokyo, 3-8-1 Komaba, Meguro-ku, Tokyo, 153-8902, Japan}

\address[KEK]{High Energy Accelerator Research Organization (KEK),
 1-1 Oho, Tsukuba, Ibaraki, 305-0801, Japan}

\begin{abstract}

Interference between different energy eigenstates in a quantum system
results in an oscillation
with a frequency which is proportional to the difference in energy
between the states.
Such an oscillation is observable in polarized positronium when it is placed 
in a magnetic field.
In order to measure the hyperfine splitting of positronium,
we perform the precise measurement of this oscillation 
using a high quality superconducting magnet and fast photon-detectors.
A result of $203.324 \pm 0.039\rm{~(stat.)} \pm 0.015\rm{(~sys.)}$~GHz is obtained
which is consistent with both theoretical calculations and previous
precise measurements.
\end{abstract}


\end{frontmatter}



\section{Introduction}
\label{intro}

Positronium~(Ps), the bound state of an electron and a positron,
is the lightest hydrogen-like atom. Since it is a purely leptonic
system and is thus free from the uncertainties of hadronic interactions,
it is an excellent object for studying Quantum Electrodynamics~(QED),
especially for the bound state.
The two ground states of positronium, the triplet state~($1^{3}S_{1}$) and 
the singlet state~($1^{1}S_{0}$), are known as 
\textit{ortho}-positronium~(\textit{o}-Ps) and \textit{para}-positronium~(\textit{p}-Ps), respectively.
The difference in the energy between \textit{o}-Ps and p-Ps is called
hyperfine splitting~(HFS), whose value is about 203~GHz,
which is significantly larger than that for the hydrogen atom~(1.4~GHz).
A theoretical prediction including $O(\alpha^3)$ corrections has been
recently obtained
using a Non-Relativistic Quantum Electrodynamics~(NRQED) approach~\cite{theory}. 
The result of this calculation deviates from the previously measured
values~\cite{MillsJr,Ritter} by a significant margin
(3.9~$\sigma$,~15~ppm).
This discrepancy might indicate the signal of new physics beyond the Standard Model similar to the discrepancy in the measurement of the muon anomalous magnetic moment~\cite{Muon_g-2}.
Both experimental results and the QED prediction should be examined again to 
confirm the discrepancy
\footnote{In the previous experimental results, there may be a systematic error in the material effect
due to non-thermalized positronium~\cite{psthermal} like
the \textit{o}-Ps lifetime puzzle~\cite{asailife}.} .

In a static magnetic field, the two states
$|s=1,m_z=0\rangle$ and $|s=0,m_z=0\rangle$ mix to 
the states $|+\rangle$ and $|-\rangle$~\cite{baryshevsky}:
\begin{eqnarray}\label{braketdef}
|+\rangle &=& C^1_1 |s=1,m_z=0\rangle + C^1_0 |s=0,m_z=0\rangle \, ,\\
|-\rangle &=& C^0_1 |s=1,m_z=0\rangle + C^0_0 |s=0,m_z=0\rangle \, ,
\end{eqnarray}
and
\begin{eqnarray}\label{constdef}
C^0_1 = -C^1_0 &=& \left\{ \frac{1}{2} \left[ 1-(1 + \chi^2)^{-\frac{1}{2}} \right] \right\}^\frac{1}{2} \, , \\
C^1_1 = C^0_0 &=& \left\{ \frac{1}{2} \left[ 1+(1 + \chi^2)^{-\frac{1}{2}} \right] \right\}^\frac{1}{2} \, ,
\end{eqnarray}
where $\chi=\frac{2g' \mu_B H}{h \Delta_{\rm{HFS}}}$, 
$\Delta_{\rm{HFS}}$ is the HFS value,
$H$ is the static magnetic field strength,
$\mu_B$ is the Bohr magneton, 
$h$ is the planck constant,
and $\varg'=\varg\left(1-\frac{5}{24}\alpha^2\right)$ is the $\varg$-factor of the electron~(positron) including the bound state correction~\cite{gcorrection}.

The other states $|s=1,m_z=\pm1\rangle$ do not couple with
the static magnetic field, and thus remain unperturbed.
The energy splitting between $|+\rangle$ and $|s=1,m_z=\pm1\rangle$~(the Zeeman splitting) is  
\begin{equation}\label{fitfunc}
\Delta_{\rm{mix}} = \frac{\Omega}{2\pi} = \frac{\Delta_{\rm{HFS}}}{2} ( \sqrt{1+\chi^2} -1 ) \, .
\end{equation}


All previous experiments obtained the value of the HFS
via the formula above
by measuring the Zeeman splitting in a magnetic field of a known
strength~\cite{asai}.
There are two distinct approaches for measuring the Zeeman splitting.
The first approach, which was proposed in Ref.~\cite{Halpern}, uses an external high power light 
source with a resonant frequency of $\Delta_{\rm{mix}}$~(about 3~GHz
in a magnetic field of $\approx$~0.8~T)
to stimulate the transition from $|s=1,m_z=\pm1\rangle$ to $|+\rangle$.
In this case, the stability of the power, frequency of the light source, and the quality of the RF cavity 
are crucial.
This approach has been used in many previous experiments, for example
Mills~{\it et al.}~\cite{MillsJr} and Ritter~{\it et al.}~\cite{Ritter}, 
and has resulted in measurements with accuracies of $O(1)$~ppm.


The second approach proposed by
V.~G.~Baryshevsky~{\it et al.}~\cite{baryshevsky} makes use 
of the quantum oscillation between $|s=1,m_z=\pm1\rangle$ and
$|+\rangle$.
Positrons emitted from a $\beta^+$ source are polarized in the
direction of their momentum
due to parity violation in the weak interaction \footnote{The polarization ratio $P$ is determined by the initial velocity
$v$ of the positron,~$P=v/c$.}.
Consequently, the resulting \textit{o}-Ps is also highly polarized.
This \textit{o}-Ps is in a superimposed state of $|+\rangle$ and $|s=1,m_s=\pm1\rangle$,
and thus oscillates with a frequency $\Omega/2\pi$.
%
%
This oscillation was observed by Baryshevsky~{\it et al.} in a subsequent experiment~\cite{baryshevskyexp}. 
This approach 
is free from systematic errors due to the high power light source and the RF cavity with a high Q-value.
Instead, a high-performance time-to-digital converter~(TDC) is crucial for precise measurement of the time spectrum.  
Those two complementary approaches are necessary to understand the discrepancy in the theoretical prediction and the experimental values of the HFS.
In 1996, S.~Fan~{\it et al.}~\cite{fan} performed an improved experiment using 
this quantum oscillation method, 
obtaining a result of $202.5 \pm 3.5$~GHz. 

In this paper, we greatly improve the accuracy of the
measurement with 
the quantum oscillation method by using 
a very high quality magnetic field, a fast photon-detection system, and high quality TDCs.
This method is based on the spin rotation of \textit{o}-Ps~(Ps-SR).
It is interesting to note that, using the relaxation of \textit{o}-Ps spin, Ps-SR 
can be used for probing various materials in material science
research~\cite{pssr}.
Since positronium is much lighter than a muon, 
the relaxation processes of positronium spin are expected to be much more 
sensitive to local magnetic structures than $\mu$-SR
\footnote{Unfortunately, the lifetime of \textit{o}-Ps is much shorter than that of a
 muon.}.


\section{Experimental setup}
\label{exp}

\begin{figure}[h]
\includegraphics[width=85mm]{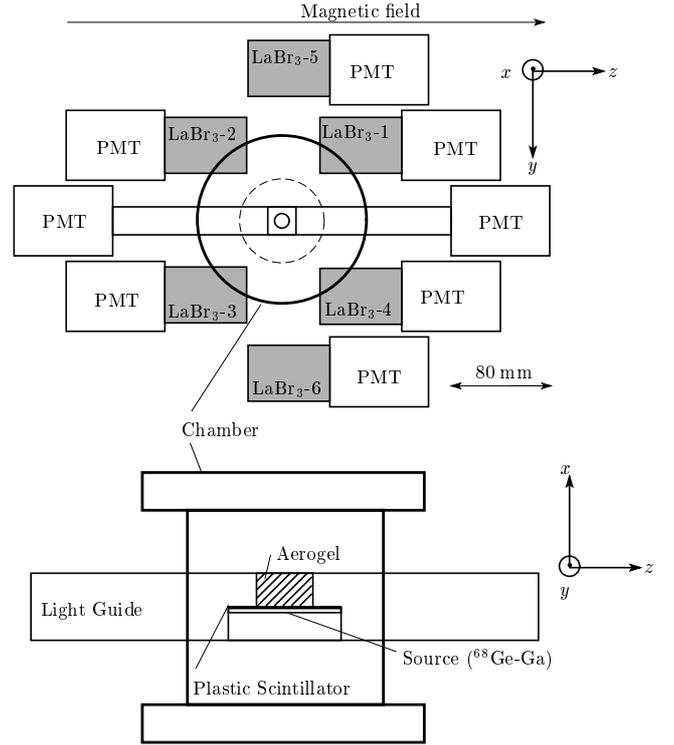}
\caption{Schematic diagram of the experimental setup. 
The top panel shows the entire experimental setup.
The magnetic field direction is along the $z$-axis,
while the LaBr$_3$(Ce) scintillators are placed in the $yz$-plane. 
The direction of the $\beta^+$ emitted
from the  $^{68}$Ge-Ga source is along the $x$-axis.
The bold circle is the positronium chamber.
The bottom panel is a magnified view of the positronium chamber, in which the
$^{68}$Ge-Ga source, the thin plastic scintillator, and the silica aerogel are located.
The coordinate systems are separately defined in the two panels.
\label{fig:setup}}
\end{figure}

The top panel of Fig.~\ref{fig:setup} shows 
a schematic diagram of the experimental setup, while the bottom panel
shows a magnified view of the positronium chamber. 
A $^{68}$Ge-Ga $\beta^+$ source,
whose end point energy is
1.9~MeV, is used.
The radioactivity is $30$~kBq, and is distributed in the active diameter of 9.35~mm.
A positron passes through a plastic scintillator~(NE102) with a thickness of 500~$\mu$m.
The resulting two light pulses are transmitted in both directions by light guides
to two photomultipliers~(PMT; Hamamatsu R5924-70).
The positron then stops and forms positronium in the silica aerogel target,
which is 10~mm in diameter and 10~mm long, with a density of 0.11~g/cm$^3$.
The surface of the primary grain is made hydrophobic 
in order to avoid the Stark effect from the electric dipole of the hydroxyl groups.
The plastic scintillator tags the positron emitted along
the direction of the $x$-axis, which results in polarized \textit{o}-Ps
along the $x$-axis.
The polarization ratio is estimated to be 0.23 by Geant4 simulation, 
in which the geometry, the threshold of the plastic scintillator and the velocity distribution of positrons are considered.
The entire positronium system is contained within a chamber evacuated 
with a rotary pump
in order to reduce pick-off annihilation~\cite{pslife}.

The magnetic field~($z$-direction) is provided by a superconducting
magnet which was originally developed for medical NMR use.
It has a large bore diameter of 80~cm and an excellent uniformity of 10~ppm 
over the volume of the silica aerogel.
The magnetic field is measured with an NMR magnetometer
~(ECHO-ELECTRONICS, EFM-150HM-AX) which has a calibration uncertainty
of 35~ppm.

The produced \textit{o}-Ps decays into three gamma rays, 
which are then detected by six LaBr$_3$ crystals
of 1.5~inches in diameter and 2~inches long. 
PMTs~(Hamamatsu R5924-70) are attached to these crystals.
The LaBr$_3$ detectors are located at 
$\left( \theta , \phi \right) = \left( \frac{\pi}{4},-\frac{\pi}{2} \right),
\left( \frac{3\pi}{4},-\frac{\pi}{2} \right),
\left( \frac{3\pi}{4},\frac{\pi}{2} \right),
\left( \frac{\pi}{4},\frac{\pi}{2} \right),
\left( \frac{\pi}{2},-\frac{\pi}{2} \right),
\left( \frac{\pi}{2},\frac{\pi}{2} \right)$,
where 
$\theta=\arccos(\frac{z}{\sqrt{x^2+y^2}})$ and
$\phi=\arctan(\frac{y}{x})$.
The detectors are referred to 
as LaBr$_3$~-~$1$ to LaBr$_3$~-~6 (see Fig.~\ref{fig:setup}). 
The quantum oscillation modulates the angular distribution of the
three gamma rays emitted from the \textit{o}-Ps decay,
which leads to beats in the decay curve of \textit{o}-Ps.
Unlike the muon precession, in which the emission direction of $\mu \rightarrow e$ rotates,
this oscillation changes its angular distribution as a \textit{vibration} in the $yz$-plane.
This is a unique property of a spin-$1$ system.
The LaBr$_3$~-~1 and LaBr$_3$~-~3 detector pair observes the oscillation with the same phase,
while the LaBr$_3$~-~2 and LaBr$_3$~-~4 detector pair observes the inverse phase. 
The LaBr$_3$~-~5 and LaBr$_3$~-~6 detectors observe the exponential decay curve without the oscillation.

\begin{figure}
\includegraphics[width=85mm]{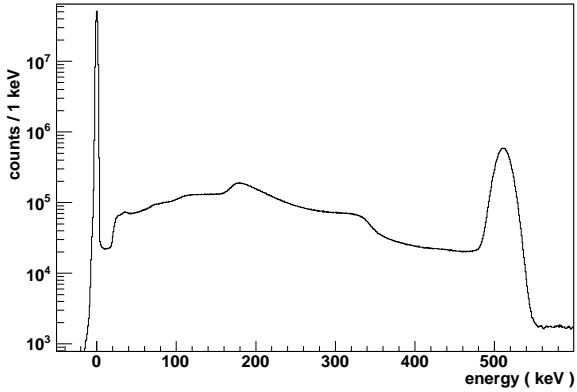}
\caption{Energy spectrum of a LaBr$_3$ crystal measured in the magnetic field of 100~mT.\label{fig:energy}}
\end{figure}

Fig.~\ref{fig:energy} shows the energy spectrum measured with one of the
LaBr$_3$ crystals.
We note that a good energy resolution of 4.0~\%~(FWHM) 
at 511~keV is obtained.
The time resolutions of the LaBr$_3$ detectors for the 511~keV gamma peak
and the positron tagging plastic scintillator are 
200~ps~(FWHM) and 3.8~ns~(FWHM), respectively.
These values are obtained with the PMTs located in a magnetic field of 100~mT.

Data acquisition starts with the rate of 740~Hz when the plastic scintillator
signal is coincident within -50~ns to 1650~ns with at least one of the
LaBr$_3$ signals. 
The time $t=0$ is defined as the timing of the plastic scintillator pulse. 
A charge ADC~(CAEN C1205) is used to measure the energy information 
of the LaBr$_3$ crystals, while 
another charge ADC~(REPIC RPC-022) is used to measure both the 
baseline information of the LaBr$_3$ crystals and the energy information of the plastic scintillator.
The baseline information is measured just before the gamma ray arrives at the LaBr$_3$ crystals 
in order to remove pile-up events.
The time differences between the plastic scintillator and LaBr$_3$ scintillators 
are measured by direct clock TDCs~(5~GHz:~time resolution of 200~ps).
These TDCs have excellent integral and differential linearities.

Separate measurements have been made for 5 different magnetic field strengths:
0~mT, 100~mT, 118~mT, 135~mT and 138~mT.  
Also, both $+x$~(up) and $-x$~(down) polarization measurements were
performed
by turning the positronium chamber upside down.
The expected time periods of the oscillation are about 26~ns and 14~ns for 
100~mT and 138~mT, respectively. 
The period of each run was about 3 days.
In total, $1.4\times10^9$ events were recorded in about 22 days of data acquisition period.
The energy and timing spectra are calibrated 
every one hour using the $511$~keV and pedestal peaks.


\section{Analysis}
\label{ana}

The following event selections are applied in order to obtain a clean time spectrum:
\begin{enumerate}
\item In order to remove pile-up events, the fluctuation of the baseline of the LaBr$_3$ is required 
to be smaller than $3\sigma$~(where $\sigma$ is the noise level).
\item The events in which more than two LaBr$_3$ crystals are hit
simultaneously are disregarded.
This helps to reduce the accidental contribution since accidental
events have a back-to-back topology.
\item In order to obtain a good time resolution, the energy deposited
in the LaBr$_3$ is required to be larger than 100~keV.
\end{enumerate}
Since the multiple hit events are removed,
six statistically independent time spectra are obtained.
We then fit the spectra using two distinct methods: 
\textit{the separate fitting method} and \textit{the subtracting method}.

\subsection{Separate fitting method}

\begin{figure}[t]
\begin{center}
\includegraphics[width=85mm]{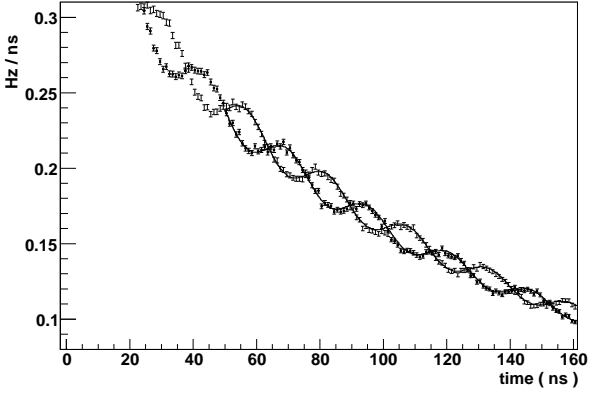}
\end{center}
\begin{center}
\includegraphics[width=85mm]{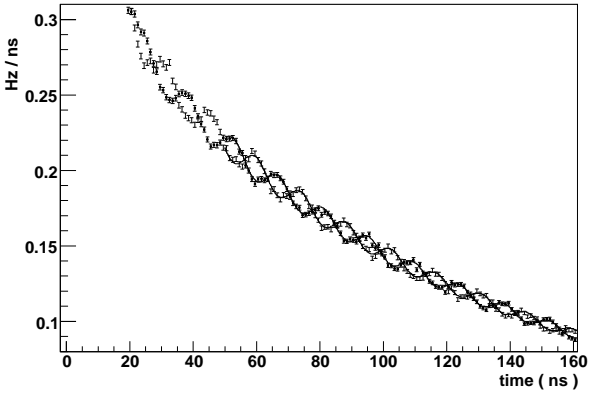}
\caption{
The timing spectra obtained in \textit{the separate fitting method} for 100~mT~(top) and 135~mT~(bottom).
In both panels, the data points are plotted with error bars, while the solid lines 
show the best fitted results.
The opposite phase spectra are superimposed in both panels.
The polarization direction of $\beta^+$ is upwards. 
\label{fig:normalfit}}
\end{center}
\end{figure}

Fig.~\ref{fig:normalfit} shows the timing spectra with the best fitted
result using \textit{the separate fitting method}.
In this method, the six time spectra are simultaneously fitted in 
the range 50~ns to 1450~ns with the following functions:
\begin{eqnarray}
f_n(t)&=&A_n e^{-\gamma_1 t}  
    +B_n e^{-\gamma_2 t} \nonumber \\
    &+&C_n e^{-\frac{\gamma_1+\gamma_2}{2} t }
    \times \sin (\Omega t + \theta_n) \nonumber \\
    &+&D_n \\&& {\rm{ for \, }} n=1,2,3,4 \nonumber\\
f_n(t)&=&A_n e^{-\gamma_1 t}  
    +B_n e^{-\gamma_2 t} \nonumber \\
    &+&D_n \\ && {\rm{ for \, }} n=5,6 \nonumber
\end{eqnarray}
where $A$ and $B$ are proportionality constants for the decay curves, 
$C$ is the oscillation amplitude, and $D$ is a constant for accidental hits
($n$ denotes the index of the LaBr$_3$ detector).
The two decay rates $\gamma_1$ and $\gamma_2$, and 
the angular frequency of the oscillation $\Omega$ are common variables among the six functions, 
while the others are kept free.
$\gamma_1$ and $\gamma_2$ are decay rates for $|s=1,m_z=\pm1\rangle$
and $|+\rangle$, respectively.
These rates include the effect of pick-off annihilation.
$\Omega$ is the Zeeman splitting frequency defined in Eq.~(\ref{fitfunc}).

\begin{figure}[t]
\begin{minipage}{1.0\hsize}
\begin{center}
\includegraphics[width=85mm]{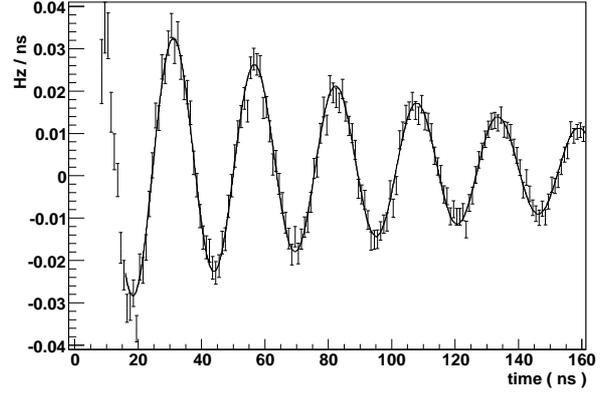}
\end{center}
\end{minipage}
\begin{minipage}{1.0\hsize}
\begin{center}
\includegraphics[width=85mm]{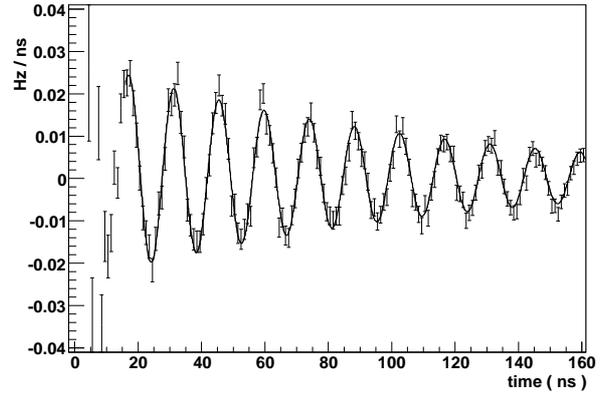}
\caption{Time spectra obtained in \textit{the subtracting method} for 100~mT~(top) and 135~mT~(bottom).
In both panels, the data points are plotted with error bars, while the solid lines 
show the best fitted results.\label{fig:subfit}}
\end{center}
\end{minipage}
\end{figure}
The results of the fits for LaBr$_3$~-~1 are listed in Table~\ref{table:fitting_example}
for the 100~mT case.
All fitted variables converged as shown in the table, in which a
reasonable $\chi^2/{\rm ndf}$ of 1.00 is obtained.
The  fitted lifetime values $1/\gamma_{1}$, $1/\gamma_{2}$ are 
136.4$\pm$2.2~ns and 102.5$\pm$2.5~ns, which are consistent
with the lifetime~($|s=1,m_z=\pm1\rangle$) measured 
in aerogel~\cite{pslife} and the
calculated value~($|+\rangle$) in a magnetic field of 100~mT,
respectively.  
A fitted time period~$2\pi/\Omega$ of 
$25.57\pm0.02$~ns~(700~ppm) is obtained for this run.

\begin{table}
\begin{minipage}{1.0\hsize}
\caption{An example of the fitted result using \textit{the separate fitting method} for 100~mT~(up).}
\begin{center}
\begin{tabular}{cc}
Parameter&Fitted Value \\
\hline
$A_1$ & $0.095\pm0.011$ \\
$B_1$ & $0.078\pm0.012$ \\
$C_1$ & $0.0096\pm0.0003$ \\
$D_1$ & $0.00733\pm0.00001$ \\
$\theta_1$~(rad) & $0.18\pm0.03$ \\  \hline
$\gamma_1$~(ns$^{-1}$) & $0.00733\pm0.00012$ \\
$\gamma_2$~(ns$^{-1}$) & $0.00975\pm0.00024$ \\
$\Omega$~(rad/ns)   & $0.24573\pm0.00017$ \\ \hline
$\chi^2/{\rm ndf}$ & 1.00~(${\rm ndf}$=8370) \\ \hline
\end{tabular}
\label{table:fitting_example}
\end{center}
\end{minipage}
\end{table}

The same procedure is applied for the runs with different magnetic
field strengths and different polarizations.
The resulting $\chi^2/{\rm ndf}$ values are always found to be less than 1.03.
$\Delta_{\rm HFS}$ can be calculated from the fitted $\Omega$.
The obtained $\Delta_{\rm HFS}$ are listed in Table~\ref{table:summerize1} 
for the various magnetic field strengths.

The 118~mT~(up) measurement is also performed 
using a different TDC clock of 8~GHz, in which consistent results are obtained.
This is an important check for the TDC, since the time spectrum 
is crucial for this experiment. 

The $\Delta_{\rm HFS}$ values obtained at the various magnetic field
strengths are consistent with each other.
The combined value is $203.336\pm0.048$~(stat.)~GHz. 
 
\begin{table*}
\begin{minipage}{1.0\hsize}
\begin{center}
\caption{
  Summary of the HFS values obtained using \textit{the separate
  fitting method}. 
(up) and (down) denote the direction of the $\beta^+$ 
with respect to the $x$-axis.
(TDC) denotes the run in which a different TDC clock was used.
Runs in the magnetic field of 118~mT have larger errors
because the runs are performed for shorter periods.
}
\begin{tabular}{ccccc}
Run&Magnetic Field~(mT) & $\Omega$~(rad/ns) & HFS~(GHz) & Events \\
\hline
100mT~(down)	&	100.592 	& 	$0.24588 \pm	0.00015$	&	$203.04 \pm 0.12$	&	$2.7\times10^{8}$	\\
100mT~(up)	&	100.594 	& 	$0.24573 \pm	0.00017$ &	$203.17 \pm 0.14$	&	$2.1\times10^{8}$	\\
118mT~(down)	&	118.824 	&	$0.34242 \pm	0.00032$	&	$203.42 \pm 0.19$	&	$4.9\times10^{7}$	\\
118mT~(up)	&	118.826 	&	$0.34289	\pm  	0.00036$	&	$203.14 \pm 0.21$	&	$6.5\times10^{7}$	\\
118mT~(up,TDC)	&	118.826 	&	$0.34207	\pm	0.00039$	&	$203.63 \pm 0.23$	&	$6.5\times10^{7}$	\\
135mT~(down)	&	134.805 	&	$0.44034	\pm	0.00025$ &	$203.58 \pm 0.11$	&	$1.9\times10^{8}$	\\
135mT~(up)	&	134.807 	&	$0.44104	\pm 	0.00030$	&	$203.26 \pm 0.14$	&	$2.1\times10^{8}$	\\
138mT~(down)	&	138.326 	&	$0.46394	\pm	0.00031$	&	$203.45 \pm 0.14$	&	$9.7\times10^{7}$	\\
138mT~(up)	&	138.330		&	$0.46414	\pm	0.00027$	&	$203.37 \pm 0.12$	&	$2.0\times10^{8}$	\\
\end{tabular}
\label{table:summerize1}
\end{center}
\end{minipage}
\end{table*}

\subsection{Subtracting method}

In this method, the sum of the LaBr$_3$~-~2 and LaBr$_3$~-~4 spectra are
subtracted from the sum of the LaBr$_3$~-~1 and LaBr$_3$~-~3 spectra.
Ideally, this would cancel the exponential components in the spectra,
leaving only the oscillating component.
However, the acceptances of the LaBr$_3$ detectors are not
exactly the same, which results in small exponential components
remaining after the subtraction.
Although these components are thus still included in the fits,
 the upshot is that the oscillating component is greatly enhanced.
Furthermore, the cancellation of the prompt peak means that the
fitting region can be extended closer to zero.
Therefore, this method has a smaller statistical error in the fits.
Fig.~\ref{fig:subfit} shows examples of the subtracted time spectra
with the best fits superimposed.
The fitting region is set from 16~ns to 1416~ns.
The fitting function is given by

\begin{eqnarray}
f(t)&=&A e^{-\gamma_1 t}  
    +B e^{-\gamma_2 t} \nonumber \\
    &+&C e^{-\frac{\gamma_1+\gamma_2}{2} t }
    \times \sin (\Omega t + \theta) \nonumber \\
    &+&D \, ,
\end{eqnarray}
where $\gamma_1$, $\gamma_2$, and $\Omega$ are the same as the ones in the separate fitting method.
The two exponential components with proportionality constants $A$ and $B$
are for the remnant decay curves, while the component with constant $C$ is
the oscillation contribution. 
The amplitudes $A$, $B$ and $D$ are expected to be small.

The fitted results are listed in Table~\ref{table:fitting_example2}
for the 100~mT case.
The coefficients $A$ and $B$ are consistent with zero and $D$ is also much smaller than $C$,
which means that the cancellation works well. 
A fitted time period~$2\pi/\Omega$ of 
$25.59\pm0.01$~ns~(590~ppm) is obtained for this run.

\begin{table}
\begin{minipage}{1.0\hsize}
\caption{An example of the fitted result using \textit{the subtracting method} for 100~mT~(up).}
\begin{center}
\begin{tabular}{cc}
Parameter&Fitted Value \\
\hline
$A$ & $0.0056\pm 0.0042$\\
$B$ & $-0.0015\pm 0.0043$\\
$C$ & $0.0377\pm 0.0008$\\
$D$ & $0.00013\pm0.00002$ \\
$\gamma_1$~(ns$^{-1}$) & $0.0077\pm 0.0010$\\
$\gamma_2$~(ns$^{-1}$) & $0.0094\pm 0.0010$\\
$\Omega$~(rad/ns)   & $0.24558\pm 0.00014$\\ 
$\theta$~(rad) & $0.16\pm 0.02$\\  \hline
$\chi^2/{\rm ndf}$ & 1.00~(${\rm ndf}$=1392) \\ \hline
\end{tabular}
\label{table:fitting_example2}
\end{center}
\end{minipage}
\end{table}

The obtained $\Delta_{\rm HFS}$ are listed in Table~\ref{table:summerize2} 
for the various magnetic field strengths.
The $\Delta_{\rm HFS}$ values obtained at the various magnetic field
strengths are consistent with each other.
The combined value is $203.324\pm0.039$~(stat.)~GHz. 

\begin{table*}
\begin{minipage}{1.0\hsize}
\begin{center}
\caption{
Summary of the HFS values obtained using \textit{the subtracting
method}. 
(up) and (down) denote the direction of the $\beta^+$ 
with respect to the $x$-axis.
(TDC) denotes the run in which a different TDC clock was used.
Runs in the magnetic field of 118~mT have larger errors
because the runs are performed for shorter periods.
}
\begin{tabular}{ccccc}
Run&Magnetic Field~(mT) &$\Omega$~(rad/ns)& HFS~(GHz) & Events\\
\hline
100mT~(down)	&	100.592  	&	$0.24579	\pm	0.00012$	&	$203.11 \pm 0.10$	&	$2.7\times10^{8}$		\\ 
100mT~(up) 	&	100.594 	&	$0.24558	\pm	0.00014$	&	$203.30 \pm 0.12$	&	$2.1\times10^{8}$		\\
118mT~(down)	&	118.824 	&	$0.34248	\pm	0.00027$	&	$203.38 \pm 0.16$	&	$4.9\times10^{7}$		\\
118mT~(up)	&	118.826 	&	$0.34289	\pm	0.00028$	&	$203.15 \pm 0.17$	&	$6.5\times10^{7}$		\\
118mT~(up,TDC)	&	118.826 	&	$0.34221	\pm	0.00031$	&	$203.55 \pm 0.18$	&	$6.5\times10^{7}$		\\
135mT~(down)	&	134.805 	&	$0.44061	\pm	0.00020$	&	$203.46 \pm 0.09$	&	$1.9\times10^{8}$		\\
135mT~(up)	&	134.807 	&	$0.44103	\pm	0.00023$	&	$203.27 \pm 0.11$	&	$2.1\times10^{8}$		\\
138mT~(down)	&	138.326 	&	$0.46401	\pm	0.00024$	&	$203.42 \pm 0.11$	&	$9.7\times10^{7}$		\\
138mT~(up)	&	138.330 	&	$0.46422	\pm	0.00022$	&	$203.34 \pm 0.09$	&	$2.0\times10^{8}$		\\
\end{tabular}
\label{table:summerize2}
\end{center}
\end{minipage}
\end{table*}


\section{Discussion and result}
\label{dis}

\subsection{Systematic errors}

The systematic errors are summarized below:
\begin{enumerate}

\item Varying the frequency sweep range of the NMR magnetometer
      results in slightly different readings. The uncertainty in the
      magnetometer calibration is estimated from this deviation~(35~ppm).

\item The non-uniformity of the magnetic field results in the following two effects: 
      (1) The oscillations in the time spectra become smeared.
       This effect is already taken into account in the fitted results listed in Table~2 and 4.
      (2) There may be a difference between the
      value of the magnetic field strength as measured by the NMR 
      and the actual values over the range of the aerogel.
          This effect is estimated at 10~ppm.

\item The accuracy of the TDC is determined by that of the clock~(Hittite HMC-T2000), 
      which is better than 2~ppm. The effects of differential and integral non-linearities
      in the TDC are negligible. 

\item The fitting region dependence is negligible as long as the
      fitting start time is later than 50~ns for \textit{the separate
      fitting method},
      and 16~ns for \textit{the subtracting method}.

\item 	In \textit{the subtracting method}, remnant exponential components could affect the HFS value 
	in fitting procedures.
	This amount is estimated by altering the scale factor for each histogram by 5\%,
	which result in the shift of 10~ppm in the HFS value.
	In \textit{the subtracting method}, weighted average of the remnant exponential components over all runs is only 0.7\%,
	which causes a negligible effect on the HFS value.

\item 	An electric field produced by aerogel grains varies the overlap of the wave function of positronium,
	which results in a shift in the HFS~(Stark effect).
	Since a hydrophobic silica aerogel~(hydroxyl groups are replaced with tri-methyl-silyl groups) is used in this experiment, 
	the effect of Stark shift is rather small. 
	The amount of Stark effect is estimated as follow~\cite{Kataoka}:
	
	The area density of silanol groups of the aerogel ($\sigma=0.44{\rm ~nm}^{-2}$) is measured in Ref.~\cite{Kataoka}.
	Although the density of aerogel is different, the primary grain in the both aerogels is the same.
	Furthermore, the same chemical processes to replace silanol group is applied.
	The average electric field which affects positronium during its flight can be calculated,
	\begin{eqnarray}
	\overline{|E|^2}&\simeq& \left( \frac{1}{4\pi\epsilon_0} \right)^2 \frac{\pi \sigma p^2}{2 \overline{L} a_{\rm Ps}^3}\\
	&=&1.0\times10^{16} \mbox{~V$^2$/m$^2$},
	\end{eqnarray}
	where $p=1.7\times10^{-18}$~esu$\cdot$cm is electric dipole moment of hydroxyl groups~\cite{dipole},
	$\overline{L}=130$~nm is the mean distance between the grains,
	and $a_{\rm Ps}=0.106$~nm is the Bohr radius of positronium.
	The amount of the shift in the HFS is then~\cite{stark}
	\begin{eqnarray}
	\frac{\delta ~ \mbox{HFS}}{\rm HFS}&=&-248\cdot\frac{\overline{|E|^2}}{E_0^2}\\
	&=& -10{\rm ~ppm},
	\end{eqnarray}
	where $E_0=m^2_{\rm e} e^5 / \hbar^4 = 5.14\times10^9$~V/cm.
	This model reproduces the experimental values in Ref.~\cite{Yam} within a factor 2.
	This Stark effect is treated as the systematic error in this analysis.

\end{enumerate}

Since $\Delta_{\rm{HFS}}$ depends on the magnetic field squared, 
the systematic errors due to the magnetic field uncertainty are doubled.
The systematic errors are combined by summing in quadrature. 

\subsection{Discussion}   

$\Delta_{\rm HFS} = $ \\
$ 203.336 \pm 0.048{\rm ~(stat.)} \pm 0.015{\rm ~(sys.)~GHz~ 
(\textit{separate fitting})}$ \\ 
$ 203.324 \pm 0.039{\rm ~(stat.)} \pm 0.015{\rm ~(sys.)~GHz~
(\textit{subtracting})}$ \\

We note that the results of the two different fitting methods are
consistent.
We use the more accurate \textit{subtracting method} value as the final
result.
The accuracy is 200~ppm, which is an improvement by a factor 90 over the
previous experiment which used the oscillation method~\cite{fan}.
This result is consistent with both the theoretical calculation~\cite{theory} 
and the previous more precise experimental values which directly
measure the Zeeman transition~\cite{MillsJr,Ritter}.

In order to observe the relaxation of positronium spin~(Ps-SR),
the oscillation amplitude is fitted as a function of time.
The result is, however, consistent with a constant.  
A better accuracy and a higher density target are necessary to
observe the relaxation.

The accuracy of the measured HFS value in this study is
200~ppm.
The following five points can be improved in order to achieve a better
accuracy: 
\begin{enumerate}
\item Increase the total run time by a factor 20~(about 1.5~years).
\item Increase the intensity of the radioactive source by a factor 3.
The dead time of the DAQ would still be acceptable.
\item Increase the coverage of the photon detectors by a factor 3.
 Fine segmentation is still necessary.
\item The absolute calibration of the NMR and the uniformity of the 
magnetic field can both easily be improved  to $O(1)$~ppm.
\item The Stark effect can be estimated by changing the density of the aerogel.
\end{enumerate}

The result of these improvements would be an increase in statistics by a factor 180
and an improvement of the final accuracy to about 15~ppm. 

\section{Acknowledgements}   

We thank Dr.~M.~Ikeno~(KEK) for the development of the TDC.
Sincere gratitude is also expressed to Mr. M.~M.~Hashimoto 
and Dr.~T.~Tanabe for the useful discussions.






\bibliographystyle{model3-num-names}
\bibliography{<your-bib-database>}







\end{document}